\documentclass[useAMS,usenatbib]{mn2e}
\usepackage{amsfonts}
\usepackage{amsmath}
\usepackage{amssymb}
\usepackage{slashbox}

\def\bea{\begin{eqnarray}}
\def\ena{\end{eqnarray}}

\newcommand{\mr}[1]{\mathrm{#1}}

\usepackage[dvips]{graphicx}
\usepackage{amsmath,amssymb}
\usepackage{myaasmacros}
\usepackage{color}
\usepackage{multirow}

\title[Fermi bubbles around M31]{Evidence of Fermi bubbles around M31}

\author[]{}

 \author[M.~S.~Pshirkov, V.~V.~Vasiliev \& K.~A.~Postnov]{M.~S.~Pshirkov$^{1,2,3}$\thanks{E-mail:
 pshirkov@sai.msu.ru }, V.~V.~Vasiliev$^{4}$\thanks{E-mail:
 vasilyev@mpia.de }, K.~A.~Postnov$^{1,5}$\footnotemark[1]\thanks{E-mail:
 pk@sai.msu.ru}, \\
$^{1}$Sternberg Astronomical Institute, 
Lomonosov Moscow State University, 
Universitetsky prospekt 13, 119992, Moscow, Russia\\
$^{2}$Institute for Nuclear Research of the Russian Academy of
Sciences, 117312, Moscow, Russia\\
 $^{3}$Pushchino Radio Astronomy Observatory, 142290 Pushchino,
Russia\\
$^{4}$IMPRS Max Planck Institute for Astronomy,D-69117, Heidelberg,	
Germany\\
$^{5}$Faculty of Physics, 
Lomonosov Moscow State University, 117234, Moscow, Russia
 }

\begin{document}

\date{}

\pubyear{2015}

\maketitle

\label{firstpage}

\begin{abstract}
Gamma-ray haloes can exist around galaxies due to the interaction of escaping galactic cosmic rays with the surrounding gas. We have searched for such a halo around the nearby giant spiral  Andromeda galaxy M31  using almost 7 years of Fermi LAT data at energies above 300 MeV. The presence of a diffuse gamma-ray halo with total photon  flux  $2.6\pm 0.6\times 10^{-9}$ cm$^{-2}$~s$^{-1}$, corresponding to a luminosity (0.3-100 GeV) of $(3.2\pm 0.6)\times 10^{38}$ erg s$^{-1}$ (for a  distance of 780~kpc) 
 was found at a 5.3$\sigma$ confidence level. The halo form does not correspond to the extended baryonic HI disc of M31, as would be expected in hadronic production of gamma photons from cosmic ray interaction, nor it is  spherically symmetric, as could be in the case of dark matter annihilation. The best-fit halo template corresponds to two 6-7.5 kpc bubbles symmetrically located perpendicular to the M31 galactic disc, similar to the 'Fermi bubbles'  found around the Milky Way centre, which suggests the past activity of the central supermassive black hole or a star-formation burst in M31.  
\end{abstract}




\section{Introduction}
\label{sec:intro}

Extended baryonic haloes around  
spiral galaxies can exist due to gas inflow from their neighbourhood
\citep{White1978,Fukugita2006}. When falling towards the galaxy, this gas can be heated up to virial temperatures $10^6-10^7$ K, producing huge 
reservoirs of hot gas (coronae). There are several observational manifestations of these coronae: soft 
diffuse X-ray emission extending up to several ten kpc from the central galaxy 
\citep{Li2008}, absorption in O VII line \citep{Wang2005,Bregman2007}, 
distortions in the shape of gas clouds \citep{Westmeier2005} and stripping of 
gas in the satellite galaxies by the ram-pressure of the halo gas 
\citep{Blitz2000}, see \citep{Putman2012} and references therein for a review. Such a hot halo around the Milky Way is 
established by several different methods \citep{Miller2013}. 


The Milky Way and other disc galaxies can also be immersed into extended cosmic-ray (CR) halos \citep{DePaolis1999,Feldmann2013}. 
It is well known that the Milky Way is not a perfect calorimeter for proton CRs: they rather quickly, on time scales of 10-20 Myr, escape from dense regions of the Galaxy, losing only minor part of their energy in interactions with the 
interstellar medium \citep{Strong2007, Strong2010}. However, if strong enough 
magnetic fields exist far away from  central regions of the Galaxy, these CRs 
would not escape to the intergalactic space, but would be instead retained in the magnetized Galactic halo for a considerable time. Magnetic fields  10-100 times as weak as the Galactic ones  
$\mathcal{O}(\mu$G) could be sufficient to contain these CRs for the cosmological time.
Wandering CRs would interact with tenuous ($\sim10^{-4}~ \mathrm{cm^{-3}}$) hot 
plasma producing gamma-rays via pionic channel. Estimates  show that the gamma-ray luminosity of such a  halo could be around $10^{39}~\mathrm{erg~s^{-1}}$ 
at energies above 100 MeV \citep{Feldmann2013}. The size and shape of the halo cannot be firmly 
established and depend crucially on the propagation properties of CRs. The halo 
'half-light' radius is estimated to be 20-40 kpc \citep{Feldmann2013}. At smaller scales ($\sim10$~kpc), the gamma-ray halo can be non-uniform as evidenced by the 'Fermi bubbles' (FB) in the Galaxy \citep{Su2010,Fermi2014}.

The contribution of the CR halo around our Galaxy to the isotropic gamma-ray background can be as high as 10$\%$, and it is difficult to disentangle it from 
the truly extragalactic component. However, such halos can be searched for around other spiral galaxies. The most natural target is the halo 
around the nearby M31 (Andromeda) galaxy. With the expected angular size of several degrees and a gamma-ray luminosity of  
$\sim10^{39}~\mathrm{erg~s^{-1}}$, such a halo could be detected by the Fermi LAT even from the Earth-M31 distance of $>700$ kpc. The presence of a hot gas around M31, which is essential for the gamma-ray emission from the CR halo, was recently demonstrated by the discovery of certain absorption features in UV-spectrum of quasars projected on the sky close to the galaxy \citep{Rao2013,Lehner2014} and distortions in the observed CMB spectrum in the vicinity of M31 due to interference from the halo gas \citep{DePaolis2014}.

%
%

\section{Data and data analysis}
\label{sec:data}

In our analysis we have used 81 months of Fermi LAT data collected  since 2008 Aug 04 ( MET =239557417 s)  
until 2015 Jul 06 (MET=457860004 s).  We have selected events that belong to the "SOURCE" 
class in order to have a sufficient number of events without loss in their 
quality. The   PASS8\_V2 reconstruction and v10r0p5\footnote{http://fermi.gsfc.nasa.gov/ssc/data/analysis/software/} version of the Fermi 
science tools  was used.
As the expected signal is weak and diffuse, we have selected events 
with energies larger than 300 MeV, because at lower energies the Fermi LAT point 
spread function (PSF) quickly deteriorates. 
Usual event quality cut, namely that the zenith angle should be   less than 
$100^{\circ}$ (which is sufficient at these energies)   has been 
imposed.

Smaller PSF allowed us to use smaller region of interest (RoI) as well -- we took a circle of 10 degrees around the centre of the M31 galaxy 
($\alpha_{J2000}=10.6846^{\circ}, \delta_{J2000}=41.2692^{\circ}$).  The data were analysed using the binned maximum likelihood approach \citep{Mattox1996} implemented in the \textit{gtlike} utility, in which two  model hypotheses were compared by  their maximal   likelihoods with respect to  the observed photon distribution. The null hypothesis   does not include the halo,  the alternative hypothesis adds the halo to the list of sources of the null hypothesis. 

The  model includes 25  sources found within RoI from the 3FGL catalogue \citep{3FGL}, the latest galactic interstellar
emission model gll$\_$iem$\_$v06$\_$rev1.fit, and the isotropic spectral template iso$\_$source$\_$v06.txt\footnote{http://fermi.gsfc.nasa.gov/ssc/data/\\access/lat/BackgroundModels.html}. Parameters (normalized flux and photon spectral index)  of 16 out of 25 point-like  and background sources were allowed to change\footnote{The significance ($TS$) of 9 out of 25 3FGL sources is less than 25, therefore their parameters were kept fixed.}. We also included additional 69 point-like gamma-ray emitters from the 3FGL catalogue found between $10^{\circ}$ and $15^{\circ}$ from the RoI centre with their parameters  held fixed.

The  M31 galaxy itself was modelled as an extended source based on the IR observations 
\citep{iras} ($100\mu$m normalized IRIS map from the InfraRed Astronomical Satellite, IRAS) following the prescriptions of the Fermi LAT collaboration 
\citep{m31_fermi}.

Finally, extended halo spatial templates were inserted into the source model. We 
have used the simplest spatial models -- uniformly bright circles of 
different radii (from 0.1$^\circ$ to 5.0$^\circ$ with 0.1$^\circ$ step). Of course, it is not a 
realistic model, because some decrease in surface  brightness towards the
outer halo regions can be expected. On the other hand, scarcity of the data used 
justifies this simple approach -- a more sophisticated  model would inevitably involve a 
 larger number of parameters, which would make fitting much harder and 
would dilute any obtained significance as well.

The M31 galaxy and the halo spectra were described by a simple power-law model:
\begin{equation}
dN/dE \propto (E/E_{0})^{-\Gamma}
\end{equation}
The normalization  and spectral index $\Gamma$ were allowed to vary during the likelihood optimisation, while the energy scale $E_{0}$ was fixed at 1 GeV. 

The evidence of detection of gamma-ray signal from the halo was evaluated in terms of a likelihood ratio test statistic:
\begin{equation}
TS = - 2 \ln \frac {L_{\mathrm{max},0}} {L_{\mathrm{max},1}}
\end{equation}
where $L_{\mathrm{max},0}$ and $L_{\mathrm{max},1}$ are maximum likelihood values obtained from the observed data fit using null and alternative hypothesis, respectively. If the \textit{alternative} hypothesis is true, then $\sqrt{TS}$ is approximately equivalent to the source detection significance.

\section{Results}
\subsection{Uniform circle template}
Firstly, we searched for diffuse gamma-ray emission from M31 galaxy itself. The  galaxy was modelled in two different ways: as a point-like source or as an extended object (the IRAS template). 
The extended template for the M31 galaxy fits the data considerably better than the simple point-like source ($TS_{\mr{ext}}=79$, $TS_{\mr{ps}}=62.3$). The galaxy has a soft spectrum with photon index $\Gamma=2.40\pm0.12$ and the  flux $F=(2.6\pm0.4)\times10^{-9} \mathrm{ph ~cm^{-2}~ s^{-1}}$ in the 0.3-100 GeV energy range. The spectrum is even softer if the galaxy is modelled as a point-like source: $\Gamma=2.64\pm0.15$ with the  photon flux $F=(1.9\pm0.3)\times10^{-9} \mathrm{ph ~cm^{-2}~ s^{-1}}$.
\begin{figure}
\begin{center}
\includegraphics[scale=0.3, angle=270]{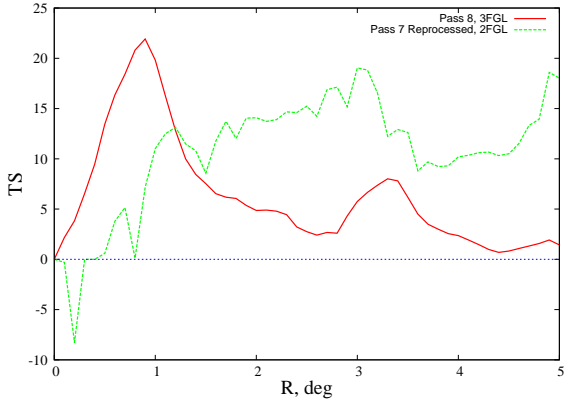}
\end{center}
\caption{ $TS (R_{\mr{halo}})$ curves: an earlier version (65 month of data, Pass7 Reprocessed events and 2FGL catalogue) is shown for  comparison. The $TS (R_{\mr{halo}})$ curve is much smoother when the latest version of event reconstruction and the  3FGL source catalogue are used.}
\label{fig:graph}
\end{figure}

\begin{figure}
\centering
{\includegraphics[width=.8\columnwidth]{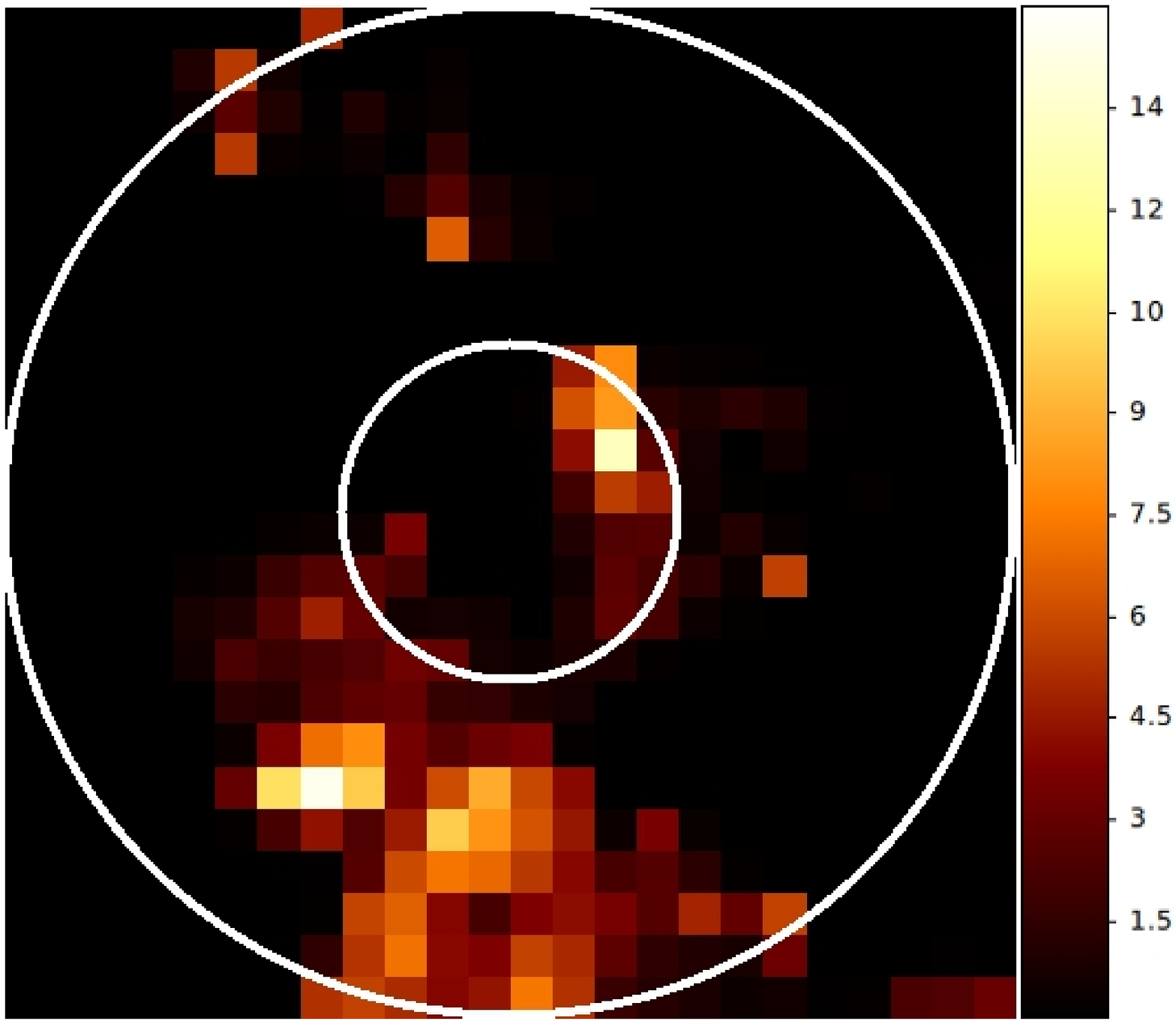}}
\vfill
{\includegraphics[width=.8\columnwidth]{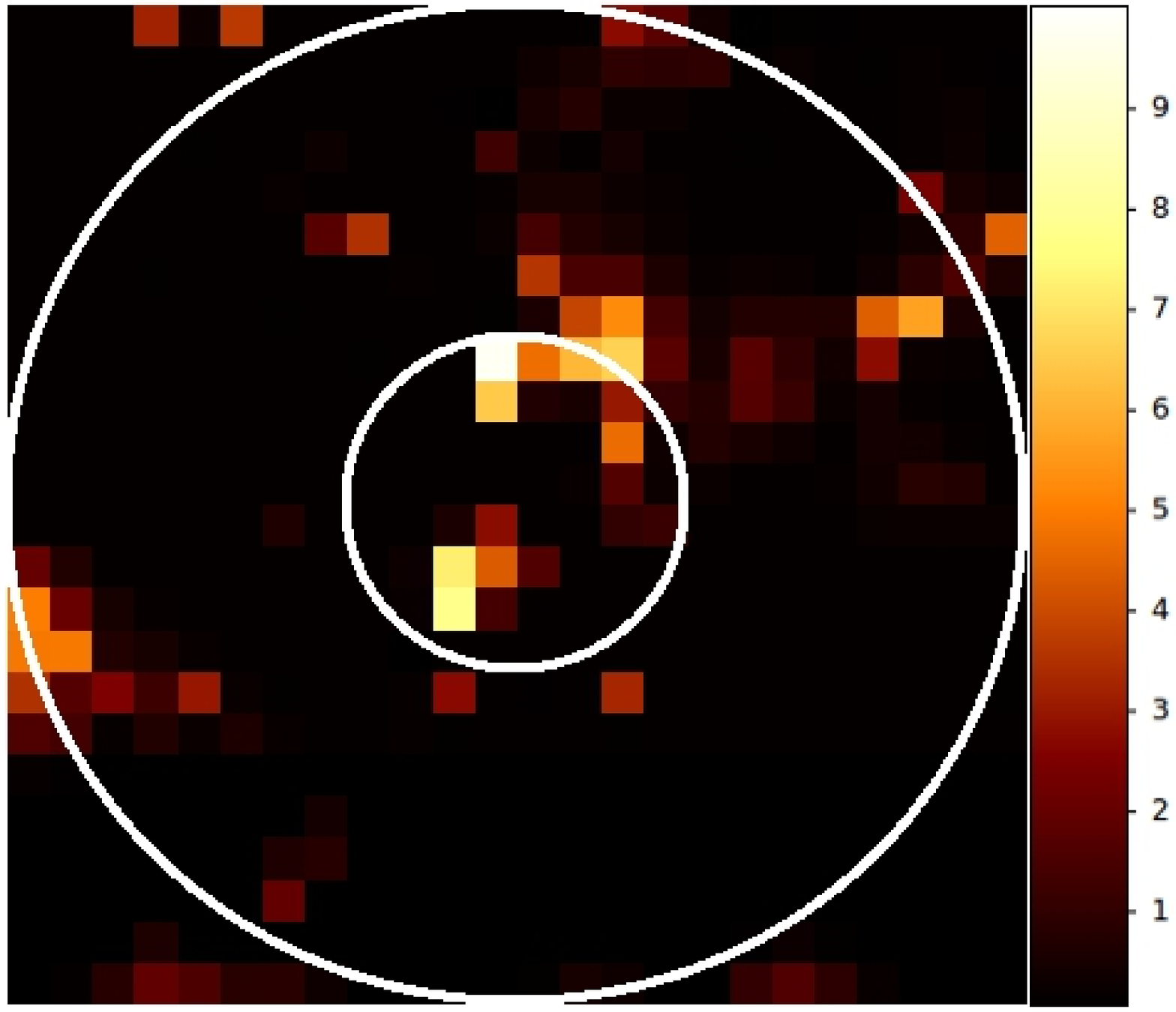}}
\caption{\textit{Upper panel:} The grey-scale  $TS$ map with the IRAS template for the M31 disc. A complex extended structure around the galaxy is clearly seen. $1^{\circ}$  and $3^{\circ}$  white circles are shown for convenience.
\textit{Lower panel:} The grey-scale  $TS$ map of simulated data including 0.9$^{\circ}$ halo.  A bright spot with $TS\sim10$ that emerged by chance is seen. There are  no clear signs of any extended structure beyond the $1^{\circ}$ radius.}
\label{fig:tsmap}
\end{figure}

 \begin{figure}
\begin{center}
\includegraphics[width=0.3\textwidth]{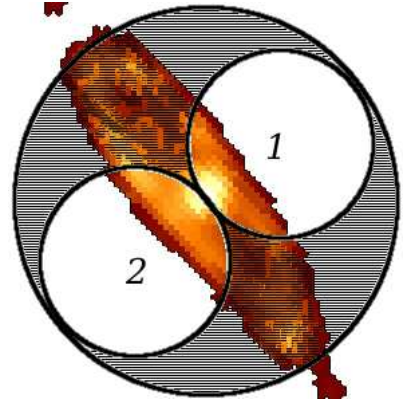}
\end{center}
\caption{ Model templates used for M31 halo fitting: IRAS template, 0.9$^{\circ}$ uniform circle, (1) and (2) -- 0.45$^{\circ}$ bubbles perpendicular to the galactic plane, and region complementary to two bubbles (the shaded area).}
\label{fig:template}
\end{figure}

 The results of fitting with additional halo component are presented in  Fig.~\ref{fig:graph}. The fit 
 quality improvement can be easily seen.
The highest statistical significance $TS=22$ was obtained for a halo with radius $R_{\mr{halo}}=0.9^{\circ}$, 
corresponding to a linear size of $\sim 12$ kpc.
The photon flux from the extended halo and the 0.3-100 GeV luminosity obtained from the fit are $\sim (3.2\pm 1.0)\times 10^{-9}$ $\mathrm{cm^{-2}s^{-1}}$ and $(4.0\pm1.5)\times 10^{38}~\mathrm{erg~s^{-1}}$, respectively, adopting the distance $d=780$~kpc. The spectral index is found to be rather soft: $\Gamma=2.30\pm0.12$.  A marginal improvement ($
TS \sim 8$)  could be also achieved by adding a 3$^\circ$ halo ($\sim35~$ kpc).

 To exclude possible systematic and instrumental effects, which could affect our results, we have performed several additional tests:

 (i)  In order to check whether the size of the PSF, which for 68\% containment is equal to  $\sim2^{\circ}(4^{\circ})$ for front- (back-) converted events at $E=300$~MeV,  could result in an extended artefact,  we  performed   Monte Carlo simulations. We have simulated events in the energy range  0.3--100 GeV for the relevant time span (71 months) and the RoI described above. 
 The model included the following components: the galactic and isotropic background,   point-like sources from the 3FGL catalogue,  the M31 galaxy (using the IRAS 100$\mu$m template) and an extended halo with R=0.9$^{\circ}$ characterized by the power-law spectral index $\Gamma=2.2$ and photon flux $F_{0.3-100 GeV}=1.5 \times 10^{-9}~\mathrm{cm^{-2}s^{-1}}$. Spectral parameters and photon fluxes for the point-like sources from  the 3FGL were taken from the 3FGL catalogue, the recommended values were chosen for the isotropic  and galactic background fluxes\footnote{http://fermi.gsfc.nasa.gov/ssc/data/analysis/scitools/help/\\	gtobssim.txt}.   After that the  simulated files were subjected to our  standard  analysis.  

  The resulting $TS$ for the halo was found to be $17.1$, the fitted power-law index was $\Gamma=1.96 \pm 0.09$ and the fitted photon flux $F_{0.3-100 GeV}=(1.46 \pm 0.23) \times 10^{-9}~\mathrm{cm^{-2}s^{-1}}$, the maximal $TS$ was reached around $R=0.9^{\circ}$. No $TS$ increase was found in this analysis if no events from the halo were simulated. Therefore, we can conclude that the finite size of the PSF or leakage from  an imperfect background treatment could produce spurious $TS$  larger than 20.

(ii) In order to check whether the initial $TS$ increase (Fig.~\ref{fig:graph})  can be  due to unidentified point-like sources not included into the 3FGL catalogue, we calculated the $TS$ map using the $gttsmap$ utility (see Fig.~\ref{fig:tsmap}, upper panel). A $TS$ excess at about 0.9$^{\circ}$ from the centre of the galaxy  with the galactic coordinates ($l=120.58^{\circ}, b=-21.17^{\circ} $) emerges that could be ascribed  to FSRQ B3  0045+013. However, even after adding  this source  into our  model, the $TS$ of the uniform halo decreased only from  22 to 15 (the $TS$ for this source was 11.2), thus the whole increase could not be attributed to this  source alone. Alternatively, this $TS$ excess could be produced by a small-scale inhomogeneity of the M31 halo. The plausibility of this scenario is also confirmed by the inspection of the $TS$ maps of several simulated haloes -- they are far 
from being smooth and uniform, but rather consist of several random knots that could have $TS>10$ (see Fig.~\ref{fig:tsmap}, lower panel).

(iii) We also checked that the smallness of our RoI does not considerably affect our analysis: we have performed the data analysis using a larger circle with 15$^{\circ}$ radius. The halo $TS$ value remained essentially unchanged.

Several additional tests are presented in the Appendix.

\subsection{Bubble template}

\begin{figure}
\begin{center}
\includegraphics[height=0.49\textwidth, angle=270]{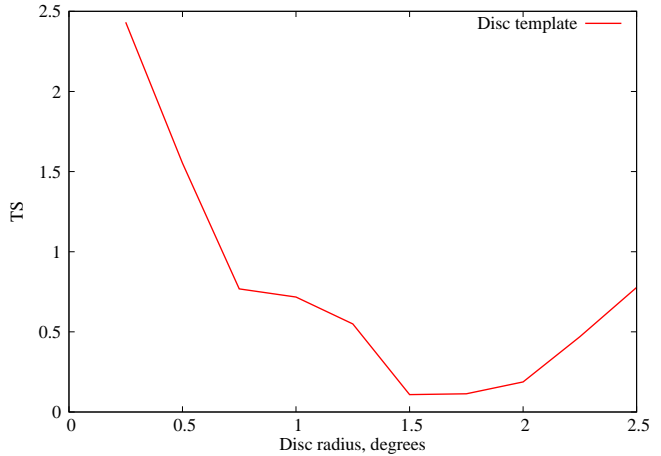}
\end{center}
\caption{ $TS$ curve for an elliptic template with  aspect ratio $a/b=0.22$, representing addition of the HI disc-like emission. Inclusion of these templates does not improve the fit quality. }
\label{fig:disc}
\end{figure}

Despite low statistics ($600-700$~photons), hints on the possible nature of the found extended emission could be inferred from its morphology. The first natural option is that the emission has hadronic origin -- the cosmic rays are interacting with the diffuse medium, producing pions and, eventually, gamma-rays. In this case the emission brightness traces the concentration of targets, i.e., the gas  density. Most of the gas mass of the M31 galaxy resides in a flat HI  disc (see, e.g \citep{Cram1980,Robles-Valdez2014}) and the expected   template    should  naturally have the shape of  an ellipse with  aspect ratio $b/a=\cos i\sim0.22$, where $a,b$  are the semi-major and semi-minor axes, respectively, and $i=77^{\circ}$ is the inclination angle. 
The results are presented in Fig.~\ref{fig:disc}: addition of such a disc-like component does not improve the fit quality. In other words, IRAS+disc template describes the data  much poorer than the  IRAS+circle one. This implies that the hadronic origin  of  the extended halo due to  the cosmic rays interactions with matter  is strongly disfavoured.

\begin{figure}
\begin{center}
\includegraphics[height=0.49\textwidth, angle=270]{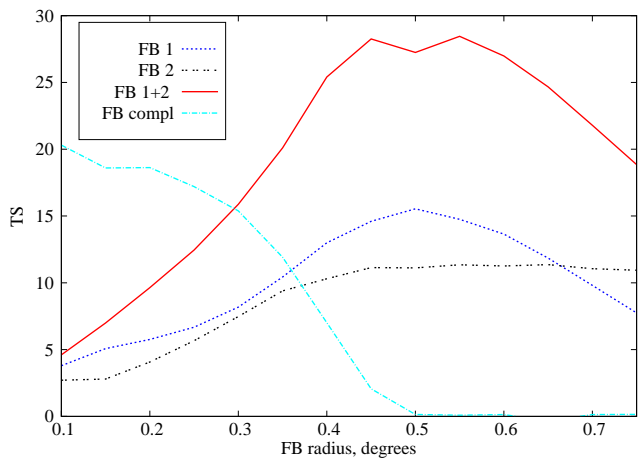}

\end{center}
\caption{ $TS$ curves for different templates. Most of the signal comes from the region corresponding to the FB 1+2 template.}
\label{fig:FB}
\end{figure}

\begin{figure}
\begin{center}
\includegraphics[height=0.49\textwidth, angle=270]{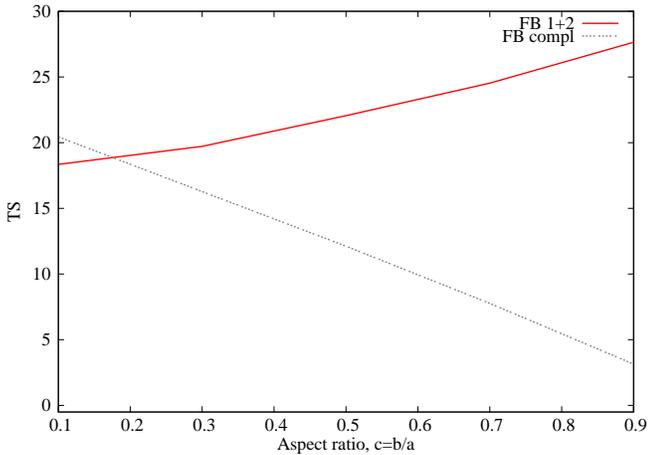}
\end{center}
\caption{ $TS$ curve for elliptic FB templates with different aspect ratio $b/a$. The semi-major axis  is fixed at $a=0.45^{\circ}$. The complementary template  is defined as above (see Fig.~\ref{fig:template})}
\label{fig:FB_ell}
\end{figure}

 The Fermi bubbles are almost circular regions of $\sim6$~kpc radius located above and below the Galactic centre, their total luminosity  is equal to $L_{\rm FB}\sim4\times10^{37}~\mathrm{erg~s^{-1}}$. They are believed to reflect past activity of the central region of the Galaxy, either from the central supermassive black hole (SMBH) or central star-formation burst. We decided to test whether a 'FB-like' morphology can fit the data better than the simple uniform disc template. 
First of all, we  performed fits varying the  bubble radius  from 0.1 to 0.75 degrees. For each bubble radius, four  different models were fitted (see Fig.~\ref{fig:template}).
The results are shown in  Fig.~\ref{fig:FB}. 
There is a clear maximum for the FB 1+2 template  with $0.45-0.55^{\circ}$ radius of the bubbles (corresponding to the linear size 6-7.5~kpc),  with $TS=28.2$. The  rest of the 0.9$^{\circ}$ circle (FB$_{\mathrm{compl}}$ template) contributes very little. The parameters of the best-fit model are listed in  Table~\ref{tab:FB}.
Additionaly, we have searched for a possible ellipticity of the bubbles. The major semi-axes of both ellipses normal to the disc plane  were fixed at $a=0.45^{\circ}$, while the aspect ratio $b/a$ was allowed to vary in the 0.1-0.9 range.  Fig.~\ref{fig:FB_ell} shows that although the signal is  concentrated around  the major semi-axes,  it is not completely contained within this narrow jet-like region.    

\begin{table}
\centering
\caption{\label{tab:FB}  Best-fit parameters for IRAS + different  halo templates from Fig.~\ref{fig:template}.  Column 1: model template, column 2: $TS$ value, column 3: integral flux in the 0.3-100 GeV range, column 4: photon spectral index $\Gamma$. For comparison, best-fit parameters for uniform 0.9$^{\circ}$ template are also presented } 
\begin{tabular}{c|c|c|c|c}
\hline
model&$TS$&\begin{tabular}{@{}c@{}}$F_{300},~10^{-9}$ \\$(\rm{cm^{-2}~s^{-1}})$\end{tabular} &$\Gamma$\\
\hline
\hline
IRAS  only& -- & $2.9\pm0.4$ &$2.4\pm0.1$ \\
FB 1 &14.6 & $1.4\pm0.5$ & $2.5\pm0.3$ \\
FB 2 &11.1 & $0.72\pm0.35$ & $2.0\pm0.2$ \\
FB 1+2 &28.2& $2.6\pm0.6$  & $2.3\pm0.1$  \\
FB$_{\mathrm{compl}}$ &2.0& $1.1\pm0.9$  & $2.5\pm0.4$  \\
\hline
0.9$^{\circ}$~circle &22.0&$3.2\pm1.0$& $2.3\pm0.1$  \\

\hline

\end{tabular}
\end{table}

\section{Discussion and conclusions}
\label{sec:conclusions}

Using almost 7 years of the Fermi-LAT observations, we  performed search for an extended gamma-ray halo at energies above 300 MeV around the closest large spiral galaxy, M31. 
We find that the Fermi-LAT data suggest the presence of a spatially extended  diffuse  gamma-ray excess around  M31.  The best-fit morphology of the diffuse emission closely resembles the Fermi bubbles in the Milky Way. The best fit gave $\sim 5.2\sigma$ significance for two $0.45^{\circ}$ (6.5 kpc) bubbles  with  a photon flux of $\sim (2.6\pm0.6)\times 10^{-9} ~\mathrm{cm^{-2}s^{-1}}$   and a  luminosity of $(3.2\pm0.6)\times 10^{38} ~\mathrm{erg~s^{-1}}$ in  the energy range 0.3--100 GeV. These parameters are fairly close to those of the Fermi bubbles in the Milky Way: $r_{\rm FB}\sim 6$~kpc,  luminosity in the 0.1-500 GeV range $4.4\times 10^{37}~\mathrm{erg~s^{-1}}$  with $\Gamma=1.9\pm0.2$ \citep{Fermi2014}. The difference in the luminosity can be ascribed to the presence of a much more massive SMBH in the M31 centre.  In view of this similarity, it would be interesting to search for a structure similar to the 'WMAP haze' at longer wavelengths \citep{Finkbeiner2004} around M31.
The bubble-like morphology of the diffuse gamma-ray emission around M31 can hardly be explained by the standard DM-related scenarios.  
Past activity of the M31 galaxy might have 
been responsible for  the  complex structure of the $TS$ excess at several degrees scale (see Fig.~\ref{fig:tsmap}, upper panel).

Our findings suggest the possible ubiquity of the FB phenomenon in giant spiral galaxies related to their central activities.  
Future observations, including at  energies $>100$ GeV \citep{VERITAS, HAWC} would certainly clarify this issue. 
 
\section*{Acknowledgements}
The work was supported by the Grant of the President of Russian Federation MK-2138.2013.2, MK-4167.2015.2 and RFBR grant 14-02-00657.  M.P. acknowledges the fellowship of the Dynasty foundation. The authors thank Drs. Dmitry Prokhorov, Igor Moskalenko, Ol'ga Sil'chenko and Anatoly Zasov for fruitful discussions. We also thank the anonymous referee for helpful suggestions. The analysis is based on data and software provided by the Fermi Science Support Center (FSSC). The numerical part of the work was done at the computer cluster of the Theoretical Division of INR RAS and cluster of the SAI MSU. 
This research has made use of NASA's Astrophysics Data System.

\bibliographystyle{mn2e}
\bibliography{m31}

\onecolumn

\section*{Appendix}
\label{Appendix}


Using \textit{gtobssim} utility, we performed several tests in order to check for possible systematics.

\begin{enumerate}
\item

First of all, as an illustration, we show that the radial profile of the observed gamma-ray excess closely matches the one expected from the M31 galaxy (simulated by the IRAS template) and 0.9$^{\circ}$ halo, see Figure~\ref{appendix:fig:radial}

\begin{figure}
\begin{center}
\includegraphics[scale=0.5, angle=270]{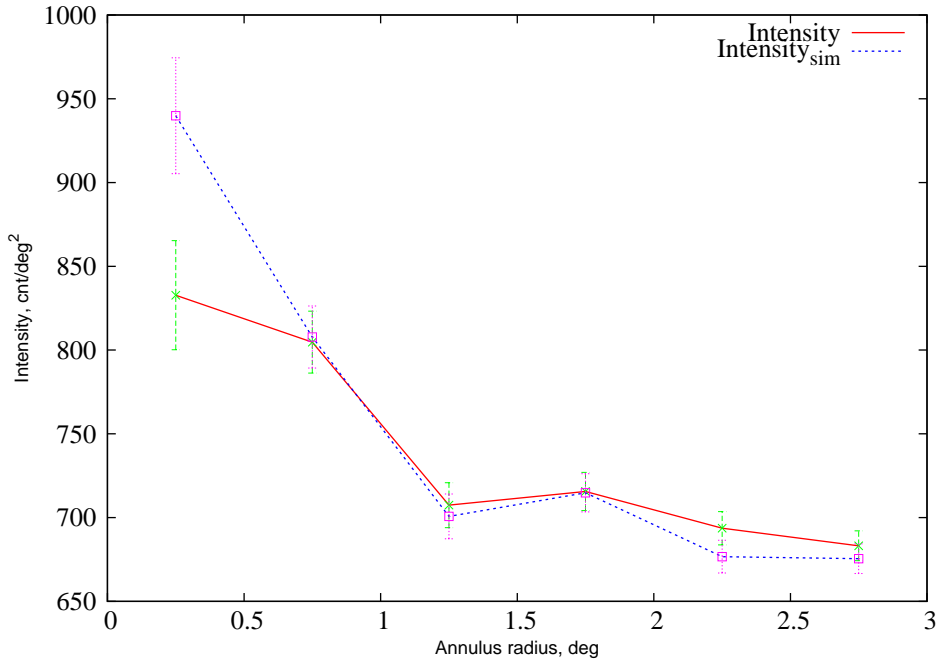}
\end{center}
\caption{The data and simulations binned in 0.5$^{\circ}$ annuli. The simulations include backgrounds, point-like sources from the 3FGL catalogue, and IRAS+0.9$^{\circ}$ halo with roughly equal contributions. The simulations well reproduce observations, being only slightly more concentrated towards the centre. This in turn may indicate that the IRAS (disc part) contribution  to the data is subdominant. 
}
\label{appendix:fig:radial}
\end{figure}

\item
We have also checked whether our method could be really sensitive to morphological features of the extended emission, since the possibility cannot be excluded that large $TS$ can be produced by chance for, e.g., FB templates, which would erroneously lead us to wrong conclusions. To check this, we simulated six sets of events with different contribution from M31:
\begin{itemize}
 \item No source;
 \item Point-like source with photon flux $F_{0.3-100 GeV}=4.0 \times 10^{-9}~\mathrm{cm^{-2}s^{-1}}$. This flux corresponds to around 1100-1200 photons and slightly overestimates the actual excess. It is used only to make the possible difference more visible. This value was adopted in all simulations;
\item IRAS template;
\item IRAS template + $0.9^{\circ}$ uniform halo with equal contributions to the total flux; 
\item $0.9^{\circ}$ uniform halo;
\item 'Fermi bubble'-like structures with $0.45^{\circ}$ radius.
\end{itemize}

These simulations were analyzed using different source models identical to ones used in the main text of the Letter:
\begin{itemize}
 \item No source;
 \item Point-like source;
\item IRAS template;
\item IRAS  + $0.9^{\circ}$ uniform halo;
\item IRAS  + FB;
\item IRAS  + FB$_\mathrm{compl}$.

\end{itemize}

The results are presented in Table~\ref{tab:dLLH}.
\begin{table}
\centering
\caption{\label{tab:dLLH} Results of the $gtlike$ fit for different simulated sets: $\Delta LLH$ for different models (with respect to the 'No source' model). Larger $\Delta LLH$ indicate better models -- our approach can distinguish different shapes of the extended emission, i.e., a uniform circle is better reconstructed with the uniform circle template, and the best reconstruction of a FB-like emission is achieved with the FB template.} 
\begin{tabular}{|c|c|c|c|c|c|c|}
\hline
\backslashbox{Simulations}{Model}& No source& Point-like & IRAS & IRAS+Halo & IRAS+FB & IRAS+ FB$_{\mathrm{compl}}$\\
\hline
\hline 
No source & 0  & 2  & 3 & -1  & - 1 & -1   \\
Point-like & 0  & \textbf{129}  & 108 & 104  &  105 & 104   \\
IRAS & 0  & 124  & \textbf{204} & 201  & 201 & 201   \\
IRAS+Halo & 0  & 102  & 172 & \textbf{172}  & 172 & 169   \\
Halo & 0  & 81  & 131 & \textbf{161}  & 152 & 142   \\
FB & 0  & 127  & 153 & 196  & \textbf{220} & 150   \\

\hline
\hline

\end{tabular}
\end{table}

\item
In the case of $0.9^{\circ}$ halo the total signal from the M31 region is dominated by the halo rather than the galaxy disc: the flux from the disc is found to be $\sim 10\%$ of the total flux $(\sim (3.3 \pm 1.0) \times 10^{-10}~\mathrm{cm^{-2}s^{-1}})$. This fact suggests that the IR-based template cannot 
fully trace the gamma-ray emission, and this emission is far more extended than the template size.  
To find how the observed gamma-ray flux from the region is shared between the two components,   
we have simulated events in the energy range  0.3--100 GeV for the relevant time span (71 months) and the RoI described above. The model included the galactic and isotropic backgrounds, 16  point-like sources from the 3FGL catalogue,  the M31 galaxy disc (taken in the form of the IRAS 100$\mu$m template). Spectral parameters and photon fluxes for the point-like sources were taken from the 3FGL catalogue. The recommended values were taken for the isotropic  and galactic backgrounds   fluxes (http://fermi.gsfc.nasa.gov/ssc/data/analysis/scitools/help/gtobssim.txt). We have fixed the total flux from the halo and disc components to a fiducial value $3.0\times	10^{-9}~\mathrm{cm^{-2}s^{-1}}$  and performed simulations, gradually  changing the halo contribution from 0 to 100 $\%$. The results are presented in Fig.~\ref{fig:flux_sim}. First of all, note that there is no leakage of the disc photons to the halo -- when the fraction of the simulated halo photons is low, the results of the corresponding fit immediately show this. The same is true for the disc component as well. Clearly, this method can  effectively separate the two components. 
      The actual flux from the M31 galaxy disc is low, $<4\times10^{-10}~\mathrm{cm^{-2}s^{-1}}$, therefore its luminosity in the 0.3-100 GeV energy range is rather modest: $5\times10^{37}~\mathrm{erg~s^{-1}}$. This value is several times as small as that of the Milky Way \citep{Strong2010}\footnote{This luminosity also includes some contribution from the Milky Way halo, so the direct comparison is not straightforward}. This difference could be explained by low level of the  star formation rate in  M31, which is smaller than the corresponding Galactic rate  by a factor of  4-5 \citep{Kennicutt2012,Ford2013}. The lower star formation rate not only implies less energy in the form  of cosmic rays that eventually produce the observed gamma-rays, but also means that the level of turbulence in the ISM of M31 is lower, which in turn leads to a rapid escape of CRs from the M31 disc region.

\begin{figure}
\begin{center}
\includegraphics[scale=0.5, angle=270]{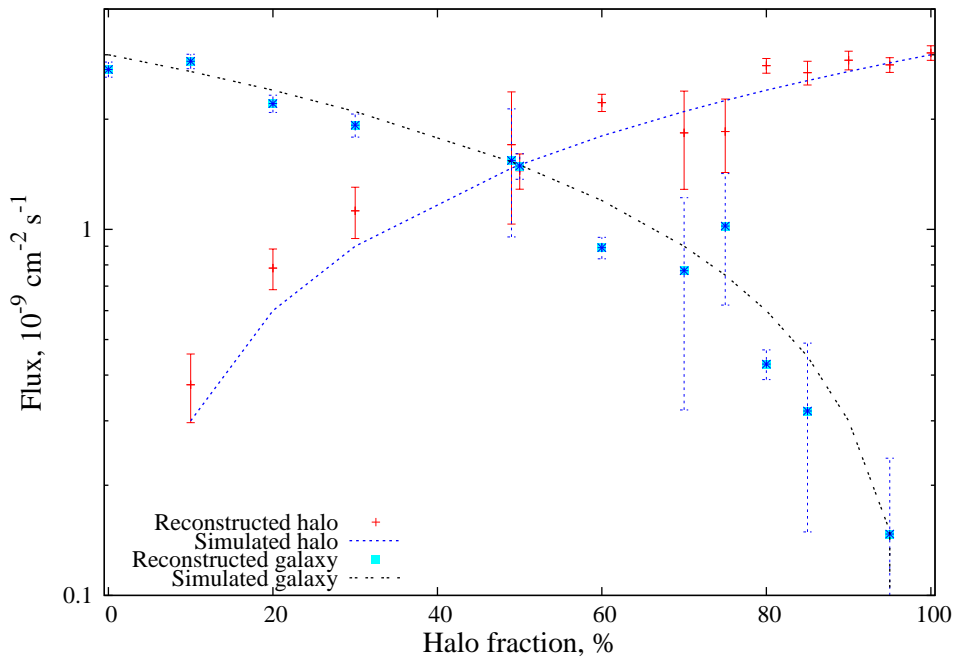}
\end{center}
\caption{ Comparison of the reconstructed and simulated fluxes from the disc and halo components for different halo fractions in the total flux. }
\label{fig:flux_sim}
\end{figure}

\item
 Finally, we show the best-fit results for all point-like sources within our RoI with parameters listed in Table~\ref{tab:fit_param}. 

\begin{table}
\centering
\caption{\label{tab:fit_param} Results of $gtlike$ fit for the  model that includes a $0.9^{\circ}$ uniform halo. Benchmark values from the 3FGL catalogue are presented for comparison.} 
\begin{tabular}{|c|c|c|c|c|c|c|}
\hline
No.& Name of the source& 3FGL normalization, & normalization    & 3FGL spectral index & Spectral index with halo & Distance, $^{\circ}$\\
 ~ & ~ &  $N_0\times 10^{14}$&  with $0.9^{\circ}$ halo, $N_0 \times 10^{14} $  ~ & ~ & ~\\
\hline\hline
1 & 3FGL J0039.1+4330 & 4.45 & 3.60$\pm$0.57 & 1.96 &  2.06$\pm$0.12 & 2.33 \\
2 & 3FGL J0040.3+4049 & 0.042 & 0.021$\pm$0.011 & 1.13 & 1.34$\pm$0.42 & 0.64 \\
3 & 3FGL J0048.0+3950 & 4.29 & 6.15$\pm$0.52 & 1.88 & 1.90$\pm$0.06 & 1.74 \\
4 & 3FGL J0049.0+4224 & 0.86 & 0.54$\pm$0.13 & 1.76 & 1.70$\pm$0.17 & 1.63 \\
5 & 3FGL J0102.3+4217 & 981 & 973$\pm$97 & 2.70 & 2.86$\pm$0.10 & 3.80 \\
6 & 3FGL J0006.4+3825 & 612 & 639$\pm$88 & 2.62 & 3.08$\pm$0.30 & 7.50 \\
7 & 3FGL J0023.5+4454 & 247 & 164$\pm$23 & 2.57 & 2.60$\pm$0.14 & 5.05 \\
8 & 3FGL J0043.8+3425 & 81.4& 134$\pm$5 & 2.04 & 1.96$\pm$0.03 & 6.84 \\
9 & 3FGL J0105.3+3928 & 59 & 62.3$\pm$6 & 2.32 & 2.17$\pm$0.08 & 4.66  \\
10 & 3FGL J0008.0+4713 & 56.8 & 54.3$\pm$2 & 2.02 & 2.00$\pm$0.03 & 5.68 \\
11 & 3FGL J0058.3+3315 & 116 & 215$\pm$147 & 2.41 & 1.39$\pm$0.62 & 8.58 \\
12 & 3FGL J0102.8+4840 & 245 & 99$\pm$99 & 1.77(PLEC) & 0.50$\pm$0.67 & 8.20 \\
13 & 3FGL J0106.5+4855 & 312 & 410$\pm$286 & 1.21(PLEC) & 1.45$\pm$0.50 & 8.72 \\
14 & 3FGL J0128.5+4430 & 60 & 3.5$\pm$11($TS \sim 2$) & 2.33 & -- & 8.98 \\
15 & 3FGL J2356.0+4037 & 1.5 & Removed  ($TS<0$) & 1.72 & -- & 8.82 \\
16 & 3FGL J2358.5+3827 & 4.8 & 361$\pm$200 & 2.07 &  1.19$\pm$0.30 & 8.93\\
\hline
\hline

\end{tabular}
\end{table}

\end{enumerate}


\end{document}